\documentclass[12pt]{article}

\input epsf
\usepackage{amssymb,wrapfig}
\usepackage{xypic,multirow}
\usepackage[matrix,arrow,curve]{xy}

\begin{document}
\addtolength{\baselineskip}{.5mm}
\newlength{\extraspace}
\setlength{\extraspace}{1.5mm}
\newlength{\extraspaces}
\setlength{\extraspaces}{2mm}

\def\nn{{\cal N}}
\def\rr {{\Bbb R}}
\def\cc {{\Bbb C}}
\def\pp {{\Bbb P}}
\def\zz {{\Bbb Z}}
\def\del {\partial}
\def\cy {Calabi--Yau}
\def\ka {K\"ahler}
\newcommand{\inv}[1]{{#1}^{-1}} 

\newcommand{\Section}[1]{\section{#1} \setcounter{equation}{0}}

\makeatletter
\@addtoreset{equation}{section}
\makeatother
\renewcommand{\theequation}{\thesection.\arabic{equation}}

\begin{center}

{ \hfill SU--ITP--06/35\\ \hfill SLAC--PUB--12266\\ \hfill UPR-1171-T}

{\Large \bf{Flux-induced isometry gauging in heterotic strings}}

\bigskip

\textbf{Wu-yen Chuang${}^{1,2}$ and Peng Gao${}^3$}

\medskip

${}^{1}${\it ITP, Stanford University, Stanford, CA 94305, USA}

${}^{2}${\it SLAC, Stanford University, Menlo Park, CA 94025, USA }

${}^3${\it  University of Pennsylvania, Philadelphia, PA 19104, USA.}

{\tt wychuang@stanford.edu, gaopeng@sas.upenn.edu}

\bigskip
{\bf Abstract}
\end{center}
We study the effect of flux-induced isometry gauging of the scalar manifold in
$N=2$ heterotic string compactification with gauge fluxes. We show that a vanishing
theorem by Witten provides the protection mechanism. The other ungauged isometries
in hyper moduli space could also be protected, depending on the gauge bundle structure.
We also discuss the related issue in IIB setting.

\section{Introduction}
It is very difficult to build a fully realistic string model without using flux compactifications \cite{DK}.
There are by now various sources of evidence suggesting that we should not restrict ourselves to the
study of Calabi-Yau spaces as string theory vacua. The study of mirror symmetry for Calabi-Yau flux compactification, for instance,
will inevitably lead us to the territory of "Non-K\"ahlerity" \cite{GM, GLMW, Fidanza, Grana, GLW,CKT}

It is also very interesting to study the fate of the well-known IIA/hetrotic string duality if we compactify
IIA string on the non-K\"ahler background. This nonpertubative duality between IIA on $K3$ fibered
Calabi-Yau and heterotic string on $K3 \times T^2$ was first studied in \cite{KLM, KV}
and then generalized to the case with fluxes and $SU(3)$-structure manifolds\cite{CKKL,LouisMicu}.
The effect of gauging induced by torsions in geometry and by various kinds of fluxes in IIA were mapped
to the gauge fluxes in heterotic string.

When we turn on the RR or NSNS fluxes in IIA/IIB/heterotic $N=2$ compactification,
supergravity analysis suggests that it will lead to the isometry gauging of the scalar manifold \cite{PS}.
This means the hypermultiplets become charged under certain vector multiplets. The gauging and the charges are specified by the killing vectors, which are determined by the fluxes turned on. The non-perturbative objects
in string theory, D-branes or D-instantons, presumably could destroy the isometries in the hyper moduli
space by introducing RR dependence into the action. In \cite{gauging}, the authors showed that the
allowed instantons in IIA string setting will not remove the flux-gauged isometries; namely the flux
will protect the gauged isometries\footnote{See \cite{alova} for a similar result in the setting of five-dimensional heterotic M-theory.}. However, other isometries are generically lifted by instanton corrections. It is not clear whether the non-perturbative correction still preserves the quaternionic structure. We notice similar arguments are not enough to reach the same conclusion in IIB case, where the shift symmetry of RR scalar $C_0$ is gauged by the NSNS flux and multiple instanton branes can contribute $C_0$ dependent corrections to the moduli space metric.

In this paper we study the $N=2$ gauged supergravity resulting from
heterotic string theory compactified on $K3 \times T^2$ with gauge
fluxes. The gauging in the supergravity analysis is achieved by
turning on the abelian gauge fluxes. The exact matching between the
IIA and heterotic flux parameters can be worked out
straightforwardly. In $N=2$ heterotic string compactification, the
hyper moduli space could receive $\alpha'$-correction \cite{aspin},
perturbatively and non-perturbatively. A worldsheet instanton
wrapping a holomorphic cycle in K3, for example, could give
correction to the hyper moduli space because there are
hypermultiplets coming from $H^2(K3)$. However, the isometry gauging
is achieved by turning on the abelian gauge fluxes over certain 2
cycle $C$ in K3 \cite{LouisMicu}, which means the gauge bundle
restricted to the 2 cycle $V|_C$ is non-trivial. This is precisely
the situation where the instanton correction is zero \cite{Witten}.

The paper is organized as follows. In section \ref{sec:IIA} we recall the isometry
protection mechanism in IIA setting. In section \ref{sec:het} we first review the
IIA/heterotic duality and then demonstrate how Witten's vanishing theorem
helps protect the gauged isometry in heterotic string. Lastly, discussion and
conclusion follow.

\section{Isometry protection in IIA flux compactificatons}
\label{sec:IIA}
In this section we begin by reviewing the isometry gauging in IIA setting and how NSNS flux
protects certain isometries \cite{gauging}. The protection follows from the tadpole consideration on the
world volume of the D-instantons.

First let us consider IIA on a Calabi-Yau $M$. Each ${N}=2$ hypermultiplet contains two complex scalars $z^a$, $a=1, \ldots, h^{2,1}$ coming from complex structure moduli of the Calabi-Yau and two scalars $\varphi^{\alpha}$, $\tilde \varphi_{\alpha}$ from expansion of the RR potential $C_3$ in a particular symplectic marking of $H^3(M)$

\begin{equation} C_3 = \varphi^{\alpha} A_{\alpha} + \tilde \varphi_{\alpha} B^{\alpha}~, \alpha=0,\ldots, h^{2,1}
\end{equation}

$\varphi^0$, $\tilde \varphi_0$, the dilaton $\phi$ and the NSNS axion $a$
form the universal hypermultiplet obtained by dualizing the tensor multiplet in four dimensions.
In the dimensionally reduced $N=2$ supergravity theory, these scalars reside on a quarternionic manifold with the metric given by \cite{ferrara1,ferrara2}:
\begin{eqnarray}
\label{eq:ihatemetric}
ds^2 &=& d \phi^2 + g_{a\bar b} dz^a d\bar z^{\bar b} + \frac{e^{4\phi}}{4}\left[da + \tilde \varphi_{\alpha} d\varphi^{\alpha} - \varphi^{\alpha} d\tilde \varphi_{\alpha} \right]\left[da + \tilde \varphi_{\alpha} d\varphi^{\alpha} - \varphi^{\alpha} d\tilde \varphi_{\alpha}\right]  \nonumber \\
& &-\frac{e^{2\phi}}{2} (\textrm{Im} \mathcal{M}^{-1})^{\alpha \beta} \left[ d\tilde \varphi_{\alpha}  +\mathcal{M}_{\alpha \gamma} d\varphi^{\gamma} \right]
   \left[d \tilde \varphi_{\beta}  +\overline{\mathcal{M}}_{\beta \delta} d\varphi^{\delta} \right]  . \nonumber
\end{eqnarray}
Expanding the background fluxes $F_4$ and $H_3$ we get
\begin{eqnarray}
F_4&=& \lambda_I \tilde{\omega}^I \nonumber\\
H _3&=& p^{\alpha} A_{\alpha} + q_{\alpha} B^{\alpha}
\end{eqnarray}
where $\tilde{\omega}^i$ a basis for $H^{2,2}(M)$.

We now have the following killing vectors corresponding to the isometries to be gauged\cite{LM, KK}:
\begin{eqnarray}
\label{eq:isometry}
(k_F)_I &=& -2 \lambda_I \partial_a  \nonumber\\
k_H {}&=& (p^{\alpha} \tilde \varphi_{\alpha} - q_{\alpha} \varphi^{\alpha}) \partial_a + p^{\alpha} \partial_{\varphi^{\alpha}} + q_{\alpha} \partial_{\tilde \varphi_{\alpha}}
\end{eqnarray}
where $F_4$ and $H_3$ determine the charges under the ${\rm I}^{th}$ vector and the graviphoton fields respectively \footnote{Throughout the paper $\alpha \beta \ldots$ and $I J \ldots$ denote hyper and vector indices respectively.}.

Due to the absence of 1 and 5-cycles in the Calabi-Yau manifold, the only relevant IIA D-instanton is the D2-instanton wrapping a 3-cycle. Consider an instanton state consisting of E2 branes wrapping a cycle in the homology class expressible as the formal sum
\begin{equation}
\label{e2config}
\Gamma_{inst} = \sum_i c^i \Gamma_i  \,.
\end{equation}
This configuration contributes a $\varphi_i$ dependence
\begin{equation}
\label{cx}
\int_\Gamma C_3 = \sum_i c^i \varphi_i
\end{equation}
to the effective action\footnote{Here we dropped the symplectic structure on $H^3$ and expand in the basis $\{\gamma^i\}$ dual to the homology basis $\{\Gamma_i\}$. We have $C_3=\varphi_i\gamma^i$ and $H_3=p_i\gamma^i$.}. Transforming the scalar manifold metric under $k_H$ we find
\begin{equation}
  \label{eq:break}
k_H (\int_\Gamma C_3) = \sum_i c^i p_i  \ .
\end{equation}
For generic values of $c^i$, the classical brane action breaks any isometry involving a shift in the value of fields $\varphi_i$.

If this were true, it will certainly destroy the consistency of the gauging procedure. However, as noticed in \cite{gauging} there is a simple mechanism at work which prohibits this from happening. The crucial observation of \cite{gauging} is that the Bianchi identity for world volume gauge flux reads
\begin{equation}
dF=-H_3~.
\end{equation}
On a compact world volume without boundary this requires
\begin{equation}
\sum_i c^i p_i =0~,
\end{equation}
from which it's obvious any physically realized instanton cannot break the gauged isometries.

A more concise way to rephrase the protection mechanism is to recall that
H flux induces magnetic charges for the brane gauge field. It implies
we can not wrap a D2-instaton over a 3 cycle on which we turned on
$H$ flux. This is simply the constraint imposed by Freed-Witten anomaly \cite{Freed}.

\section{Isometry protection in heterotic string}
\label{sec:het}
In this section we will review the IIA/heterotic duality with
gauge fluxes. We wil provide an exact matching between the
flux parameters \cite{KV, CKKL, LouisMicu}.  Then we show a theorem due to Witten guarantees the protection of gauged isometries.

\subsection{IIA/heterotic duality}
The IIA/heterotic duality was first studied in \cite{KLM, KV}. Besides the spectrum matching, the conifold
transitions in IIA string on CY is mapped to the Higgsing of the charged hypermultiplets in
heterotic string. The Higgsing can move the theory around the different moduli space strata
with different dimensions. This beautiful phenomenon is not the topic of our paper although
the transition in the presence of fluxes is certainly worth further studying.

We will begin by recalling the results in \cite{LouisMicu, CKKL}. The anomaly cancellation
in the 10d supergravity requires we modify the heterotic $H$ in the following way,
\begin{equation}
H= dB + \omega_{gravity} - \omega_{YM} .
\end{equation}

From this we get a new Bianchi identity:
\begin{equation}
dH = tr R \wedge R - tr F \wedge F
\end{equation}
where $R$ is the Riemann curvature of the internal manifold and $F$ is the field strength
of the Yang-Mills fields.

For heterotic string on $K3 \times T^2$, we will need 24 instanton number to cancel the $\int_{K3} tr( R \wedge R) $
contribution. In earlier literature, people usually studied the gauge bundle with $c_1(V)=0$. But in fact there exists
no obstruction for us to turn on $c_1$ of the gauge bundle (equivalent to turning on abelian gauge fluxes). It is also
possible to turn on $c_1$ such that it does not contribute to $\int  tr (F \wedge F)$. This can be seen as follows.
Let us first turn on the following gauge fluxes over the 2 cycles in K3,
\begin{equation}
\label{eq:gflux}
\int_{\gamma^{\alpha}} F^{I}_{gauge} = m^{\alpha I},\ \ I=0 ,\cdots, n_V, \ \ \alpha=1, \cdots, 22.
\end{equation}
where $I$ is the index for vector moduli and the zeroth component stands for the graviphton. These fluxes could contribute to the tadpole condition \cite{LouisMicu, OldLM}:
\begin{equation}
\int_{K3} tr ( F \wedge F ) + \delta = 24
\end{equation}
where
\begin{equation}
\delta=  \int_{K3} F^{I}_{gauge} \wedge F^{J}_{gauge} \eta_{IJ} = m^{\alpha I} m^{\beta J} \rho_{\alpha \beta}
\eta_{IJ}
\end{equation}

$\rho_{\alpha \beta}$ is the K3 intersection matrix with signature $(3,19)$ and $\eta_{IJ}$
is the invariant tensor on $SO(2, n_V-1)$. As we will see later, turning on $m^{\alpha I}$ in heterotic is
dual to turning on various kinds of fluxes in IIA. So if we start from IIA on K3-fibered CY with fluxes and
are interested in finding its heterotic dual with gauge fluxes, we should consider $\delta =0$ so that
the originally balanced tadpole condition will not be disturbed.\footnote{We would have to solve the
anomaly cancellation condition from the very beginning if $\delta$ does not vanish.}

Now let us consider the gauging effect of turning on gauge fluxes accroding to (\ref{eq:gflux}) over 20 2 cycles in
an attractive $K3$, following \cite{LouisMicu}. \footnote{ In \cite{CKKL}
the gauge fluxes are turned on over the $P^1$ of the $K3$ in heterotic string, which corresponds
to $F$ with support on the base of the $K3$ fibered Calabi-Yau. This flux will charge the axion in heterotic string.
The IIA dual of the gauge flux through $T^2$ fiber in heterotic K3 is unknown.}
After expanding ten-dimensional $\mathcal{B}$ filed in terms of the $H^2(K3)$ basis $\omega_{\alpha}$,

\begin{equation}
\mathcal{B} = B + b^{\alpha} \omega_{\alpha}
\end{equation}

the covariant derivative of $b^{\alpha}$  becomes

\begin{equation}
D b^{\alpha} = d b^{\alpha} - (\eta_{IJ} m^{\alpha J}) A^I = d b^{\alpha} - m_I^{\alpha} A^I
\end{equation}

The resulting killing vector is
\begin{equation}
\label{eq:gfgauge}
k_I = -  m_I^{\alpha} \partial_{b^{\alpha}}  .
\end{equation}

Recall that $b^{\alpha}$ in heterotic string corresponds to some $\varphi^{\alpha}$ in (\ref{eq:ihatemetric}).
Comparing (\ref{eq:gfgauge}) with (\ref{eq:isometry}), we immediately see that
the gauging from the gauge fluxes does not correspond to any $H$ or $F$ in
IIA theory. But the effect can be dualized into the gauging coming from the torsions
in the geometry; it is dual to IIA on an $SU(3) \times SU(3)$ structure manifold \cite{LouisMicu}.
For a review on $SU(3) \times SU(3)$ structure manifolds in the context of supergravity, see
\cite{GLW}.

Note that there are also bundle moduli coming from the gauge bundle.
Their number is determined by the dimension of the sheaf cohomology group. In this case the
sheaf will be the endormophism of the gauge bundle \cite{HLBook}.
It is easy to show that the first order deformation of this sheaf is $H^1(K3, E^{*} \otimes E )$,
where $E^{*}$ is the dual sheaf. The dimension counting, which includes the high order obstruction,
 can be done by computing the Euler character $\chi(E,E)$,

\begin{equation}
\chi(E,E) = \sum (-1)^i \textrm{dim} Ext^{i} (E,E) = \int_X ch(E^{*}) ch(E) \sqrt{Td(X)}
\end{equation}

Let now X be a K3 surface. If $E$ is a coherent sheaf on X with $rk(E) = r$, $c_1(E)= c_1$,
and $c_2(E) = c_2$, the complex dimension of the bundle moduli is given by $2 r c_2 - (r-1) c_1^2 - 2 (r^2 -1)$.

At this moment it is not clear now to charge these bundle moduli under the gauge fields because we know very little
about the hyper moduli space. We will revisit this problem in the future. In the next section, we will demonstrate the mechanism
which protects the gauged isometries from gauge fluxes in heterotic string.

\subsection{Witten's vanishing theorem}

In this section we will show how Witten's result \cite{Witten} can protect the gauging in
$N=2$ heterotic theory. In the previous section, the gauging of the $b^{\alpha}$ results
from turning on the gauge flux over the corresponding 2 cycle $\gamma^{\alpha}$. So
the worldsheet instanton wrapping $\gamma^{\alpha}$ could break this gauging, by a calculation
similar to in section \ref{sec:IIA}. Namely we can integrate $\mathcal{B}$ over $\gamma^{\alpha}$
and find that $k (\int_{\gamma^{\alpha}} \mathcal{B}) \neq 0$, where $k$ is the killing vector.

In \cite{Witten}, it was shown that the worldsheet instanton correction to the hyper moduli
space is given by

\begin{equation}
U_{\gamma^{\alpha}} = \textrm{exp} ( - \frac{A(\gamma^{\alpha})}{2 \pi \alpha'} + i \int_{\gamma^{\alpha}} \mathcal{B} ) \,
\frac{\textrm{Pfaff}  (\bar \del_{V(-1)})}{(\textrm{det}' \bar \del_{\mathcal{O}})^{4}}
\end{equation}
The exponential factor comes from the classical instanton action while the rest is the one-loop determinant from fluctuations around the classical solution. More precisely, the $\textrm{Pfaff}  (\bar \del_{V(-1)})$ in the numerator comes from one loop determinant of non-zero modes of the left-moving fermions. Three powers of $(\textrm{det}' \bar \del_{\mathcal{O}})$ come from the complex bosons representing the non-compact ${\bf{R}}^4$ directions and the $T^2$ factor. The remaining one follows by partly canceling the contribution of the normal bundle $(\textrm{det}\nabla_{\mathcal{O}(-2)})$ in $K3$ against the right moving fermions.

It is in general a very hard problem to compute this quantity. Fortunately  the theorem states that $U_{\gamma^{\alpha}}$
vanishes if and ony if the gauge bundle $V$ restricted to $\gamma^{\alpha}$ is non-trivial\footnote{This is equivalent to that the operator $(\bar \del_{V(-1)})$ has a nonempty kernel. For our purpose, $V$ can be taken as the abelian vector bundle where the gauge flux sits and $V(-1)={\mathcal{O}}(-1)\otimes V$.}. The non-trivial gauge
bundle is always the case if we want to gauge the isometry in heterotic string. It is also very likely that
the bundle restriction is already non-trivial before turning on the gauge fluxes. In this case, some ungauged isometries
are also protected \footnote{For example, we can embed the K3 spin connection into the gauge group. The bundle will be non-trivial
along every 2 cycle in K3.}. But we should keep in mind the possibility that the theory can move in the bundle moduli space
such that the bundle becomes trivial along some 2 cycle in K3 and then the worldsheet instantons re-appear.

The other potential worry is that the $U(1)$s coming from $T^2$ and
graviphoton do not belong to the $E_8 \times E_8$ bundle in the
heterotic string. Turning on their gauge fluxes do not change the
bundle restriction $V|_{\gamma^{\alpha}}$. Therefore, in order to
protect the gauged isometries, we need the bundle structure to be
non-trivial along the 2 cycles along which we turn on the gauge
fluxes. The study of the protection mechanism becomes model
dependent; we have to know the bundle structure first before
commenting on whether certain gauged isometries are lifted.

Nonetheless, in heterotic string the protection of the flux-induced
gauging is still stronger than the IIA case. In IIA case, we have no
gauging protection mechanism if we don't turn on $H$ and the
ungauged isometries are generically lifted by the quantum effects.

\section{Discussion and conclusion}

In this paper we study the flux-induced isometry gauging in $N=2$ heterotic string compactified on
$K3 \times T^2$ with gauge fluxes. A vanishing theorem by Witten \cite{Witten} guarantees that
the gauging is protected against the worldsheet instanton effect. In heterotic string, the isometry
protection can even reach the ungauged ones, which is contrary to IIA. In IIA we can not protect
the gauging without $H$ and usually lose the ungauged isometry due to D-instanton effects.
However, it is still not clear how to charge the bundle moduli under the vector moduli, which is also
contrary to IIA, where any hyper moduli can be charged under the vectors.

In the IIB case, the situation remains obscure since various branes with different dimensions come into play. Especially the $D1$ instanton wraps a 2 cycle and the Freed-Witten anomaly argument does not eliminate its existence. In view of the relation between IIB and type I theory, it seems possible that a combination of $H$ flux and the argument discussed here can achieve the protection of gauging in IIB. One could also try to study the closely related problem in $N=1$ orientifold setting. We leave this for future study.

{\bf Acknowledgments:}  We would like to thank Cingular \& Verizon Wireless for providing
phone service. WYC received support from the DOE under contract DE-AC03-76SF00515.  PG is  supported in part by DOE grant DE-FG02-95ER40893.


\end{document}